\documentclass[10pt]{iopart}
\usepackage{iopams,bm,cite,color,graphicx,mathptmx,url}

\newcommand{\op}[1]{\hat{#1}}
\newcommand{\openone}{\leavevmode\hbox{\small1\normalsize\kern-.33em1}}

\eqnobysec

\begin{document}

\title{Quantum polarization tomography of bright squeezed light}

\author{C~R~M\"{u}ller$^{1,2}$, B~Stoklasa$^{3}$, C~Gabriel$^{1,2}$, C~Peuntinger$^{1,2}$,
  J~\v{R}eh\'{a}\v{c}ek$^{3}$, Z~Hradil$^{3}$, A~B~Klimov$^{4}$,
  G~Leuchs$^{1,2}$, C~Marquardt$^{1,2}$ and  L~L~S\'anchez-Soto$^{1,2,5}$}

\address{$^{1}$ Max-Planck-Institut f\"ur die Physik des Lichts,
  G\"{u}nther-Scharowsky-Stra{\ss}e 1, Bau 24, 91058 Erlangen,
  Germany}

\address{$^{2}$ Institut f\"ur Optik, Information und Photonik,
  Staudtstra{\ss}e 7, 91058 Erlangen, Germany}

\address{$^{3}$ Department of Optics, Palacky University,
  17. listopadu 12, 771 46 Olomouc, Czech Republic}

\address{$^{4}$ Departamento de F\'{\i}sica, Universidad de
  Guadalajara, 44420~Guadalajara, Jalisco, Mexico}

\address{$^{5}$ Departamento de \'Optica, Facultad de F\'{\i}sica,
  Universidad Complutense, 28040~Madrid, Spain}

\date{\today}

\begin{abstract}
  We reconstruct the polarization sector of a bright polarization
  squeezed beam starting from a complete set of Stokes measurements.
  Given the symmetry that underlies the polarization structure of
  quantum fields, we use the unique SU(2) Wigner distribution to
  represent states. In the limit of localized and bright states,
  the Wigner function can be approximated by an inverse
  three-dimensional Radon transform. We compare this direct
  reconstruction with the results of a maximum likelihood estimation,
  finding an excellent agreement.
\end{abstract}


\section{Introduction}

Polarization of light is a robust characteristic that can be
efficiently manipulated using modest equipment without introducing
more than marginal losses. It is thus not surprising that this is
often the variable of choice to encode quantum information, as one can
convince oneself by looking at some recent cutting-edge experiments,
including quantum key distribution~\cite{Muller:1993rr}, quantum dense
coding~\cite{Mattle:1996wd}, quantum
teleportation~\cite{Bouwmeester:1997nx}, rotationally invariant
states~\cite{Radmark:2009oq}, phase
super-resolution~\cite{Resch:2007kl}, and weak
measurements~\cite{Dixon:2009hc}.

In the discrete-variable regime of single, or few, photons, one is
mostly interested into two-mode states, which for all practical
purposes can be regarded as a spin
system~\cite{Lamas-Linares:2001ul,Sehat:2005wd}. As a result, the
polarization state can be determined from correlation functions of
different orders~\cite{White:1999fk,Kwiat:2000rw,James:2001vn,
  Thew:2002pd,Barbieri:2003ij,Bogdanov:2004bs,Moreva:2006fv,
  Barbieri:2007uq,Adamson:2010ys,Sansoni:2010zr,Altepeter:2011ly}.
Given the small dimensionality of the Hilbert space involved, the
state reconstruction can be readily performed.

In the continuous-variable case, polarization properties are exploited
for an expedient generation, manipulation, and measurement of
nonclassical light.  Polarization squeezing~\cite{Chirkin:1993dz,
  Korolkova:2002fu, Luis:2006ye,Mahler:2010fk}, which has been
observed in numerous experiments~\cite{Bowen:2002kx,Heersink:2003oz,
  Dong:2007fu,Shalm:2009mi,Iskhakov:2009pi}, is perhaps the most
tantalizing illustration. Full Stokes
polarimetry~\cite{Brosseau:1998lr} is the method employed by the
majority of the practitioners in this area.

However, the reconstruction in this limit is a touchy business and
requires special care.  The origin of the problem can be traced back
to the fact that the characterization of the polarization state by the
whole density operator is superfluous, because it contains much more
than polarization information.  This redundancy can easily be handled
for low number of photons, but becomes a real hurdle for highly
excited states. An adequate solution has been proposed recently: it
suffices with a subset of the density matrix that has been called the
``polarization sector''~\cite{Raymer:1999fc,Raymer:2000zt} or the
``polarization density operator''~\cite{Karassiov:1993lq}.  Its
knowledge allows for a complete specification of the state on the
Poincar\'e sphere (actually on a set of nested spheres that can be
appropriately called the Poincar\'e space). The technique was devised
by Karassiov and coworkers~\cite{Bushev:2001xq,Karassiov:2004xw,
  Karassiov:2005ss} and implemented experimentally
in~\cite{Marquardt:2007bh}.

In this paper we present a comprehensive treatment of polarization
tomography. As with any reliable quantum tomographical scheme, we need
to supply three key ingredients~\cite{Hradil:2006os}: the availability
of a tomographically complete measurement, a suitable representation
of the quantum states, and a robust algorithm for inverting the
experimental data. In this respect, we use a standard Stokes scheme
that implements the first ingredient in a very simple way; for the
second, we resort to the well-known SU(2) Wigner
distribution~\cite{Stratonovich:1956qc,Berezin:1975mw,Agarwal:1981bd,
  Brif:1998if,Varilly:1989ud,Heiss:2000kc,Klimov:2000zv,Klimov:2008yb},
and finally, we prove that the inversion of the data in terms of that
Wigner function is an inverse three-dimensional (3D) Radon transform.

To support the experimental feasibility of our scheme, we carry out
the full tomography of a bright polarization-squeezed state generated
in a Kerr medium~\cite{Heersink:2005ul}. The reconstruction is
accomplished in three-different ways: by the direct inversion of the
Radon transform, by a maximum-likelihood estimation and, finally, by a
Gaussian approximation. The final results are compared and the sources
of uncertainty are analyzed.

\section{Setting the theoretical foundations}

\subsection{Polarization structure of quantum fields}

A satisfactory description of the polarization structure of quantum
fields and the corresponding observables that specify this structure
is of paramount importance for our purposes.

We restrict our attention to the case of a monochromatic plane wave
(the formalism can be easily extended to more involved multimode
wavefronts~\cite{Karassiov:2006hq,Karassiov:2007gb}), which we assume
to propagate in the $z$ direction, so its electric field lies in the
$xy$ plane. Under these conditions, we are dealing with a two-mode
field that can be fully characterized by two complex amplitude
operators. They are denoted by $\op{a}_{H}$ and $\op{a}_{V}$, where
the subscripts $H$ and $V$ indicate horizontally and vertically
polarized modes, respectively. The commutation relations of these
operators are
\begin{equation}
  \label{bccr}
  [\op{a}_{k}, \op{a}_{\ell}^\dagger ] = \delta_{k\ell} \, , 
  \qquad
  k, \ell \in \{H, V \} \, .
\end{equation}
The description is greatly simplified if we use the Schwinger
representation~\cite{Schwinger:1965kx,Chaturvedi:2006vn}
\begin{equation}
  \label{Stokop}
  \fl
  \op{J}_{1} = \textstyle\frac{1}{2} ( \op{a}^{\dagger}_{H}  \op{a}_{V} + 
  \op{a}^{\dagger}_{V} \op{a}_{H} ) \, ,
  \qquad
  \op{J}_{2} =  \frac{i}{2} ( \op{a}_{H} \op{a}^{\dagger}_{V} - 
  \op{a}^{\dagger}_{H} \op{a}_{V} ) \, , 
  \qquad
  \op{J}_{3}  = \frac{1}{2} ( \op{a}^{\dagger}_{H} \op{a}_{H} - 
  \op{a}^{\dagger}_{V} \op{a}_{V} ) \, ,
\end{equation}
together with the total number
\begin{equation}
  \op{N} = \op{a}^{\dagger}_{H} \op{a}_{H} + 
  \op{a}^{\dagger}_{V} \op{a}_{V}  \, .
\end{equation}
These operators coincide, up to a factor 1/2, with the Stokes
operators~\cite{Luis:2000ys}, whose average values are precisely the
classical Stokes parameters~\cite{Born:1999yq}. Using
equation~(\ref{bccr}), one immediately notices that $\mathbf{\op{J}} =
(\op{J}_{1}, \op{J}_{2}, \op{J}_{3})$ satisfy the commutation
relations of the su(2) algebra
\begin{equation}
  \label{crsu2}
  [ \op{J}_{k}, \op{J}_\ell] = i \epsilon_{k \ell m} \,  \op{J}_{m}  \, ,
\end{equation}
where $\epsilon_{k \ell m}$ is the Levi-Civita fully antisymmetric
tensor.  This noncommutability precludes the simultaneous exact
measurement of the physical quantities they represent. Among other
consequences, this implies that no field state (apart from the vacuum)
can have sharp nonfluctuating values of all the operators
$\op{\mathbf{J}}$ simultaneously. This is expressed by the uncertainty
relation
\begin{equation}
  \Delta^{2} \op{\mathbf{J}}  = 
  \Delta^{2} \op{J}_1  + \Delta^{2} \op{J}_2  + \Delta^{2} \op{J}_3  
  \geq \langle \op{N} \rangle /2 \, ,
\end{equation}
where the variances are given by $\Delta^{2} \op{J}_{i} = \langle
\op{J}_{i}^{2} \rangle - \langle \op{J}_{i} \rangle^{2}$.  In other
words, the electric vector of a monochromatic quantum field never
traces out a definite ellipse.

In classical optics, the total intensity is a well-defined quantity
and the Poincar\'e sphere appears thus as a smooth surface with radius
equal to that intensity.  In quantum optics we have
\begin{equation}
  \label{eq:nhmr}
  \op{J}_{1}^{2}  + \op{J}_{2}^{2}  +  \op{J}_{3}^{2} = 
  \left (    \frac{\op{N}}{2} \right ) 
  \left (   \frac{\op{N}}{2} +   \openone \right )  \, , 
\end{equation}
and, as fluctuations in the number of photons are unavoidable
(leaving aside photon-number states), we are forced to talk of a
three-dimensional Poincar\'e space (with axis $J_{1}$, $J_{2}$ and
$J_{3}$) that can be envisioned as foliated in a set of nested spheres
with radii proportional to the different photon numbers that
contribute significantly to the state.

The Hilbert space $\mathcal{H}$ of these two-mode fields has a
convenient orthonormal basis in the form of Fock states for both
polarization modes, namely $|n_H, n_V \rangle$. However, since
\begin{equation}
  [ \op{N}, \op{\mathbf{J}} ] = 0 \, ,
\end{equation} 
each subspace with a fixed number of photons $N$ must be handled
separately. In other words, in the previous onion-like picture of the
Poincar\'e space, each shell has to be addressed independently.  This
can be emphasized if instead of the Fock basis, we employ the
relabeling
\begin{equation}
  \label{Fockbas}
  | J, m \rangle \equiv | n_H = J + m, n_V = J - m \rangle \, ,
\end{equation}
According to (\ref{eq:nhmr}), we have that $J = N/2$ and this basis
can be also seen as the common eigenstates of $\{ \op{J}^{2},
\op{J}_{3}\}$. In this way, for each fixed $J$ (i.e., fixed number of
photons $N$), $m$ runs from $-J$ to $J$ and these states span a
$(2J+1)$-dimensional subspace wherein $\op{\mathbf{J}}$ acts in the
usual way (in units $\hbar = 1$)
\begin{eqnarray}
  & \op{J}_{\pm} |J, m \rangle  =  \sqrt{J( J + 1 ) - m (m \pm 1)}
  | J, m \pm 1 \rangle \, , & \nonumber \\
  & & \\
  & \op{J}_{3} | J, m \rangle  =  m | J, m \rangle \, , &  \nonumber
\end{eqnarray}
with $\op{J}_{\pm} = \op{J}_{1} \pm \op{J}_{2}$.

It is clear from all this previous discussion that the moments of any
energy-preserving observable (such as $\op{\mathbf{J}}$) do not depend
on the coherences between different subspaces.  The only accessible
information from any state described by the density matrix
$\op{\varrho}$ is thus its polarization sector, which is specified by
the block-diagonal form
\begin{equation}
  \op{\varrho}_{\mathrm{pol}} = \bigoplus_{J}  \hat{\varrho}^{(J)} 
\end{equation}
where $\op{\varrho}^{(J)}$ is the reduced density matrix in the $J$
subspace. Any $\op{\varrho}$ and its associated block-diagonal form $
\op{\varrho}_{\mathrm{pol}} $ cannot be distinguished in polarization
measurements (and, accordingly, we drop henceforth the subscript pol).
This is consistent with the fact that polarization and intensity are,
in principle, independent concepts: in classical optics the form of
the ellipse described by the electric field (polarization) does not
depend on its size (intensity).

\subsection{Polarization squeezing and the dark plane}
\label{sec:darkplane}

The variances of the angular-momentum operators (\ref{Stokop}) are not
independent, for they are constrained by
\begin{equation}
  \label{eq:polsquez1}
  \Delta^{2} \op{J}_{k} \,  \Delta^{2} \op{J}_{\ell}  \ge 
  \epsilon_{k\ell m} \, | \langle \op{J}_{m} \rangle  |^{2} \, .
\end{equation}
It is always possible to find pairs of maximally conjugate operators
for this uncertainty relation. This is equivalent to establishing a
basis in which only one of the operators (\ref{Stokop}) has a nonzero
expectation value, say \mbox{$\langle\op{J}_{k} \rangle = \langle
  \op{J}_{\ell}\rangle=0$} and $\langle\op{J}_{m} \rangle \neq 0$.
The only nontrivial Heisenberg inequality reads thus
\begin{equation}
  \Delta^2 \op{J}_{k} \, \Delta^2 \op{J}_{\ell}  \geq 
  |\langle\op{J}_m \rangle|^2 \, .
\end{equation}
Polarization squeezing can then be sensibly defined
as~\cite{Chirkin:1993dz,Korolkova:2002fu,Luis:2006ye,Mahler:2010fk}:
\begin{equation}
  \Delta^2 \op{J}_{k} < |\langle\op{J}_{m} \rangle| 
  < \Delta^2 \op{J}_{\ell} \, .
  \label{eq:polsq1}
\end{equation}
The choice of the conjugate operators $\{\op{J}_{k},\op{J}_{\ell} \}$
is by not means unique: there exists an infinite set  
$\{ \op{J}_\perp(\theta), \op{J}_\perp(\theta+\pi/2) \}$
that are perpendicular to the state classical excitation $\op{J}_{m}$,
for which $\langle\op{J}_\perp(\theta) \rangle =0$ for all
$\theta$. All these pairs exist in the $J_{k}$--$J_{\ell}$ plane, which
is called the ``dark plane'' because  it is the plane of zero mean
intensity. We can express a generic $\op{J}_{\perp}(\theta)$ as
$\op{J}_{\perp}(\theta) = \op{J}_{k} \, \cos \theta + \op{J}_{\ell} \,
\sin \theta$, $\theta$ being an angle defined relative to
$\op{J}_{k}$.  Condition (\ref{eq:polsq1}) is then equivalent to
\begin{equation}
  \Delta^2 \op{J}_\perp (\theta_{\mathrm{sq}}) < 
  \textstyle\frac{1}{2} | \langle \op{N} \rangle| < 
  \Delta^2 \op{J}_\perp ( \theta_{\mathrm{sq}} + \pi/2 ),
  \label{eq:polsq2}
\end{equation}
where $\op{J}_\perp(\theta_{\mathrm{sq}} )$ is the squeezed parameter
and $\op{J}_\perp( \theta_{\mathrm{sq}} + \pi/2 )$ the antisqueezed
parameter.

In the experiments presented in this paper, a focal role will be played by
the example in which the horizontal and vertical modes have the same
amplitude but are phase shifted by $\pi / 2$: $\langle \op{a}_{H}
\rangle = i \langle \op{a}_{V} \rangle = i \alpha / \sqrt{2}$,
$\alpha$ being a real number.  This light is circularly polarized and
fulfills $\langle \op{J}_{1} \rangle = \langle \op{J}_{3} \rangle =
0$, $\langle \op{J}_{2} \rangle = \alpha^{2}$. It is advantageous to
work in the circular polarization basis, whose right $(+)$ and left
$(-)$ amplitudes are given in terms of the linear ones by
\begin{equation}
  \op{a}_{\pm} =  \frac{1}{\sqrt{2}}  
  ( \op{a}_{H} \pm i \,  \op{a}_{V} ) \, .
\end{equation}
In this manner, $\langle \op{a}_{+} \rangle = \alpha$ and $\langle
\op{a}_{-} \rangle = 0$.  The operators in the $J_{1}$--$J_{3}$ plane
correspond to the quadrature operators of the dark left-polarized
mode. In fact, expressing the fluctuations of $\op{\mathbf{J}}$ in
terms of the noise of the circularly polarized modes $\delta
\op{a}_{\pm}$ and assuming $|\langle \delta \op{a}_{\pm} \rangle| \ll
\alpha$ we find~\cite{Corney:2008uq}
\begin{equation}
  \delta \op{J}_\perp (\theta) =  \alpha \delta\op{X}_{-} (\theta) = 
  \alpha [ \delta \op{X}_{H} (\theta ) + 
  \delta \op{X}_{V} (\theta +  \pi/2) ] \, , 
  \label{eq_polsq_darkmode}
\end{equation}
where $\op{X}_{i} = (\op{a}_{i} e^{-i \theta} + \op{a}_{i}^{\dagger}
e^{{i \theta}} )/\sqrt{2}$ are the rotated quadratures for the $i$th
amplitude. On the other hand, we have that
\begin{equation}
  \delta \op{N}  = 
  \alpha ( \delta\op{a}_{+} + \delta \op{a}^\dagger_{+}  )=  
  \alpha \, \delta\op{X}_{+} \, ,
  \label{eq_polsq_brightmode}
\end{equation}
and the intensity exhibits no dependence on the dark mode.  In
consequence, the condition (\ref{eq:polsq2}) can be recast for this
example as
\begin{equation}
  \Delta^{2} \op{J}_\perp (\theta) <  |\langle\alpha\rangle|^2
  \quad
  \Leftrightarrow 
  \quad 
  \Delta^2 \op{X}_{-} (\theta )  < 1 \, ,
  \label{eq_sq_equivalence}
\end{equation}
that is, polarization squeezing is equivalent to vacuum squeezing in
the orthogonal polarization mode.

In the dark-plane measurements, the beam is divided equally between
two photodetectors. Such measurements are then identical to balanced
homodyne detection: the classical excitation is a local oscillator for
the orthogonally polarized dark mode. The phase between these modes is
varied by rotating the measurement through the dark plane, allowing a
full characterization of the noise properties. This is a unique
feature of polarization measurements and has been used in many
experiments~\cite{Grangier:1987fk,Smithey:1993uq,Schnabel:2003vn,
  Julsgaard:2004kx, Josse:2004ys}.

\subsection{Polarization Wigner function}

The structure discussed so far highlights that SU(2) is the symmetry
group for polarization. To provide an appropriate phase-space
description of states, we take advantage of the pioneering work
of Stratonovich~\cite{Stratonovich:1956qc} and
Berezin~\cite{Berezin:1975mw}, who introduced quasi-probability
distribution functions on the sphere satisfying all the proper
requirements.  This construction was later generalized by
others~\cite{Agarwal:1981bd,Brif:1998if,Varilly:1989ud,Heiss:2000kc,
  Klimov:2000zv,Klimov:2008yb} and has proved to be very useful in
visualizing properties of spinlike systems~\cite{Dowling:1994sw,
  Atakishiyev:1998pr,Chumakov:1999sj,Chumakov:2000le,Klimov:2002cr}

To gain physical insights into this approach, let us start by
representing the density matrix with respect to the polarization
basis. Instead of using directly the states $\{ | J, m \rangle \}$, it
is more convenient to write such an expansion as
\begin{equation}
  \label{rho1}
  \op{\varrho}^{(J)} =  \sum_{K= 0}^{2J} \sum_{q=-K}^{K}  
  \varrho_{Kq}^{(J)} \,   \op{T}_{Kq}^{(J)} \, ,
\end{equation}
where the irreducible tensor operators $T_{Kq}^{(J)}$
are~\cite{Varshalovich:1988ct}
\begin{equation}
  \label{Tensor} 
  \op{T}_{Kq}^{(J)} = \sqrt{\frac{2 K +1}{2 J +1}} 
  \sum_{m,  m^{\prime}= -J}^{J} C_{Jm, Kq}^{Jm^{\prime}} \, 
  |  J , m^\prime \rangle \langle J, m | \, ,
\end{equation}
with $ C_{Jm, Kq}^{Jm^{\prime}}$ being the Clebsch-Gordan coefficients
that couple a spin $J$ and a spin $K$ \mbox{($0 \le K \le 2J$)} to a total
spin $J$.  These tensor operators have the right
transformation properties under rotations and they indeed constitute
the most suitable orthonormal basis
\begin{equation}
  \Tr  [ \op{T}_{K q}^{(J)} \, 
  \op{T}_{K^{\prime} q^{\prime}}^{(J^{\prime}) \, \dagger}   ] =
  \delta_{J J^{\prime}}  \delta_{K  K^{\prime}} \delta_{q q^{\prime}} \, .
\end{equation}
Although at first sight they might look a little bit intimidating,
they are nothing but the multipoles used in atomic
physics~\cite{Blum:1981ya}: one can check that
\begin{equation}
  \label{eq:multi}
  \op{T}_{00}^{( J )} = \frac{1}{\sqrt{2 J + 1}} \op{\openone} \, 
  \qquad
  \op{T}_{1q}^{( J )} = \sqrt{\frac{3}{( 2 J + 1 ) (J+1) J}} \, \op{J}_{q}
  \quad q = \pm, z \, , 
\end{equation}
and similarly the $\op{T}^{(J)}_{Kq}$ can be related to the $K$th
power of the generators (\ref{Stokop}).  Accordingly, the expansion
coefficients
\begin{equation}
  \varrho_{Kq}^{(J)} =  \Tr [ \op{\varrho}^{(J)} \,
  T_{Kq}^{(J) \, \dagger} ]
\end{equation}
are known as state multipoles.

The Wigner function associated with the state (\ref{rho1}) is
\begin{equation}
  \label{eq:WignSU2}
  W^{(J)} (\theta, \phi) = \Tr [ \op{\varrho}^{(J)} \, 
  \op{\Delta}^{(J)} (\theta,  \phi) ] \, ,
\end{equation}
where $\op{\Delta}^{(J)} (\theta, \phi) $ is the Wigner kernel
\begin{equation}
  \label{eq:WignkerSU2}
  \op{\Delta}^{(J)} (\theta, \phi ) = \sqrt{\frac{4 \pi}{2 J +1 }}
  \sum_{K=0}^{2J} \sum_{q=-K}^{K} Y_{Kq}^{\ast} (\theta, \phi) \,
  \op{T}_{Kq}^{(J)} \, ,
\end{equation}
and $Y_{Kq} (\theta, \phi)$ are the spherical harmonics.  This kernel
is unitary and satisfies the normalization conditions
\begin{equation}
  \label{eq:3}
  \Tr [ \op{\Delta}^{(j)} (\theta, \phi ) ] = 1 \, , 
  \qquad
  \frac{2 J + 1}{4 \pi} \int_{\mathcal{S}^{2}} d\Omega \ \op{\Delta}^{(j)}
  (\theta, \phi ) = \op{\openone} \, .
\end{equation}
The integral extends over the unit sphere $\mathcal{S}^{2}$ and
$d\Omega$ is the invariant measure therein, namely, $d\Omega = \sin
\theta \, d\theta d\phi$.

From (\ref{eq:WignSU2}) and the properties of the irreducible tensors,
one can immediately express the Wigner function in the very suggestive
form
\begin{equation}
  \label{eq:WignSU2Agar}
  W^{( J )} (\theta, \phi ) = 
  \sum_{K=0}^{2J} \sum_{q=-K}^{K} 
  \varrho_{Kq}^{(J)}  \; Y_{Kq}^{\ast} (\theta, \phi) \, ,
\end{equation}
which clearly shows that determining this Wigner function is
tantamount to the knowledge of all the state multipoles. (i.e., all
the moments of the Stokes parameters).
 
One can obtain the marginal of $W^{(J)}(\theta ,\phi )$ once summed
over all the values of $J$
\begin{equation}
  W (\theta ,\phi ) = \sum_{J} \frac{2J+1}{4 \pi}
  W^{(J)} (\vartheta ,\varphi ) \, ,
\end{equation}
where the factor has been introduced to ensure the proper
normalization.

In the above-mentioned example of a strong circularly polarized state,
we can consider that the sphere can locally be replaced by its tangent
plane since $J \simeq \alpha^{2}$. Using simple geometrical
relations between the coordinates $(\theta, \phi)$ and the Cartesian
coordinates $(q, p)$ in that tangent plane, we get
\begin{equation}
  \label{eq:approx}
  W (\theta ,\phi ) \simeq \alpha   \; W(q, p) \, ,
\end{equation}
which confirms that the dark plane is equivalent to the standard phase
space for continuous variables.

In the limit of large photon numbers the representation
(\ref{eq:WignSU2}) is not very useful. In such a case, a remarkably
effective approximation is given by\cite{Klimov:2000kh}
\begin{equation}
  \label{JosaA}
  \op{\Delta}^{(J)} (\theta ,\phi ) \simeq (-1)^{J }
  \exp (-i \pi \mathbf{n} \cdot \op{\mathbf{J}} ) \, , 
  \nonumber
\end{equation}
where $\mathbf{n} = (\cos \theta \sin \phi, \sin \theta \sin \phi,
\cos \theta)$ is the unitary vector in the direction $(\theta ,\phi
)$.

\subsection{Tomograms and tomographic inversion}

A general polarimetric apparatus consists of a half-wave plate, with
axis at angle $\alpha$, followed by a quarter-wave plate at angle
$\beta$. For fixed values of the angles $(\alpha, \beta)$ of the wave
plates, the selected direction in the Stokes space is
\begin{equation}
  \label{eq:pums}
  \theta = \pi/2 - 2 \beta \, ,
  \qquad
  \phi = 2 \beta - 4 \alpha \, .
\end{equation}
The polarization transformations performed by the wave plates can be
represented by $\op{J}_2$, which generates rotations about the
direction of propagation, and $\op{J}_3$, which generates phase shifts
between the modes. Their joint action is given by the operator
\begin{equation}
  \label{eq:dispsph}
  \op{D} (\theta, \phi) = e^{ i \theta \op{J}_2} \, 
  e^{ i \phi    \op{J}_3} \, ,
\end{equation}
which describes displacements over the sphere.  After that, a
polarizing beam splitter projects onto the basis $| J, m \rangle$.

In physical terms, the wave plates transform the input polarization
allowing the measurement of different Stokes parameters by the
projection onto the basis $| J, m \rangle$.  This can be modeled by
\begin{equation}
  \label{eq:POVM1}
  \op{\Pi}_{m}^{(J)} = |J, m \rangle \langle J, m | \, ,
\end{equation}
so that $w_{m}^{(J)} = \Tr [ \op{\varrho} \op{\Pi}_{m}^{(J)} ]$ is the
probability of detecting $n_H = J+m$ photons in the horizontal mode
and simultaneously $n_V= J-m$ photons in the vertical one. Of course,
when the total number of photons is not measured and only the
difference $m$ is observed, it reduces to
\begin{equation}
  \Pi_{m} = \sum_{J = |m|}^\infty |J, m \rangle \langle J, m | \, .
\end{equation}

The experimental histograms recorded for each direction $(\theta,
\phi)$ correspond to the tomographic probabilities
\begin{equation}
  \label{cojo}
  w^{(J)}_{m} (\theta, \phi ) =  
  \Tr [ \op{\varrho} \,  \op{\Pi}_m^{(J)} (\theta, \phi ) ] =  
  \Tr [ \op{\varrho} \, \op{D}  (\theta, \phi) \, \op{\Pi}_m^{(J)}
  \,  \op{D}^\dagger (\theta, \phi ) ] \, .
\end{equation}
The reconstruction in each $(2J+1)$-dimensional invariant subspace can
now be carried out exactly since it is essentially equivalent to a
spin $J$~\cite{Brif:1999kx,Amiet:1999vn,DAriano:2003ys,Klimov:2002zr}.
One can proceed in a variety of ways, but perhaps the simplest one is
to look for an integral representation of the tomograms (\ref{cojo});
as soon as one realizes that
\begin{equation}
  \label{eq:intreppovm}
  \op{\Pi}_{m}^{(J)} (\theta, \phi) = \frac{1}{2 \pi} \int_{0}^{2\pi}
  d\omega \, 
  \exp [ i \omega (\op{\mathbf{J}} \cdot \mathbf{n} - m ) ] \, ,
\end{equation}
the tomograms read as
\begin{equation}
  \label{eq:intreptom}
  w_{m}^{(J)} (\theta, \phi) = \frac{1}{2 \pi} \int_{0}^{2\pi}
  d\omega \, \Tr [ \op{\varrho}^{(J)} \, 
  \exp ( i \omega \op{\mathbf{J}} \cdot  \mathbf{n} ) ] \, e^{- i \omega  m }   \, ,
\end{equation}
that is, they appear as the Fourier transform of the characteristic
function for the observable $\op{\mathbf{J}} \cdot \mathbf{n}$. After
some direct manipulations, we find that
\begin{equation}
  \label{uf} 
  \op{\varrho}^{(J)} =  \frac{1}{4 \pi} \sum_{m=-J}^{J}
  \int_{\mathcal{S}_2} d \mathbf{n}^\prime \  
  w_m^{(J)}  (\mathbf{n}^\prime) \, 
  \mathcal{K} ( \op{\mathbf{J}} \cdot \mathbf{n}^\prime - m ) \, ,
\end{equation}
where $d \mathbf{n}^\prime$ indicates integration over the unit sphere
and the kernel $\mathcal{K} (x) $ is
\begin{equation}
  \label{kersing}
  \mathcal{K}  ( x ) = \frac{ 2 J +1}{4 \pi^2} \int_{0}^{2\pi}d\omega
  \; \sin^2 \left ( \frac{\omega}{2} \right ) \, e^{-i \omega  x} \, .
\end{equation}
Although (\ref{uf}) is a formal solution, it is handier to map this
density matrix onto the corresponding Wigner function, for which we
need to compute $ \Tr [ \mathcal{K} (\mathbf{J} \cdot
\mathbf{n}^{\prime} - m) \, \op{\Delta}^{(J) }(\theta ,\phi )]$. When
$J$ is large enough, we can replace the Wigner kernel by its
approximate expression (\ref{JosaA}), getting
\begin{equation}
  \fl 
  \Tr [ \mathcal{K} (\mathbf{J} \cdot \mathbf{n}^{\prime} -  m ) \, 
  \op{\Delta}^{(J) }(\theta ,\phi ) ] = (-1)^{J}
  \frac{2 J+1}{4\pi ^{2}} \int_{0}^{2\pi }d\omega \, 
  \sin^{2} \left ( \frac{\omega }{2} \right )
  e^{-im \omega}  \chi_{J} (\omega ^{\prime }) \, ,
\end{equation}
where $\chi _{J} (\omega ^{\prime })$ is the character of the
$(2J+1)$-dimensional representation of the SU(2)
group~\cite{Varshalovich:1988ct} and $\omega^{\prime}$ is given by
\begin{equation}
  \cos \left ( \frac{\omega^{\prime}}{2} \right ) =
  \mathbf{n} \cdot \mathbf{n}^{\prime} \, 
  \sin \left ( \frac{\omega}{2} \right ) \, .
\end{equation}
For $ J \gg~1$, $m$ can be taken as a continuous variable. Replacing
the sum by an integral, integrating by parts and taking into account
that for localized states $\mathbf{n} \cdot \mathbf{n}^{\prime} \simeq
1$, the Wigner function simplifies to
\begin{equation}
  \label{3DRT}
  W (J, \theta, \phi ) = \frac{2J+1}{4\pi^2} 
  \int_{-\infty }^{\infty} \!  dm \int_{\mathcal{S}_2} \!  
  d\mathbf{n}^\prime \, \frac{d^2 w_{m}^{(J)}  ( \mathbf{n})}{dm^2}
  \,  \delta (m - J \, \mathbf{n} \cdot \mathbf{n}^\prime ) \, ,
\end{equation}
where we have included $J$ as an argument to stress that it must be
treated as continuous.  The reconstruction in this limit turns out to
be equivalent to an inverse Radon transform of the measured tomograms.

\section{Experiment}

\subsection{The setup}

To validate our approach, we perform the tomography of a polarization
squeezed state, generated in a polarization-maintaining optical fibre
through the nonlinear Kerr effect~\cite{Heersink:2005ul}. The setup is
shown in figure~1.  The light source is a shot-noise limited ORIGAMI
laser from Onefive GmbH emitting 220~fs pulses at a repetition rate of
80~MHz and centered at 1560~nm. The light is fed into a 13~m-long
polarization-maintaining birefringent fibre (3M FS-PM-7811, 5.6 $\mu$m
mode-field diameter) so that quadrature squeezed states are
simultaneously and independently generated in both polarization modes.
The strong birefringence of the fibre (beat length 1.67~mm) causes a
delay between the emerging quadrature-squeezed pulses. We
precompensate for this delay in an unbalanced Michelson-like
interferometer placed before the fibre. A small part (0.1~\%) of the
fibre output serves as the input to a control loop to maintain the
relative phase between the exiting pulses locked to $\pi/2$, so the
light is circularly polarized.

\begin{figure}[b]
  \centering{\includegraphics[width=0.80\columnwidth]{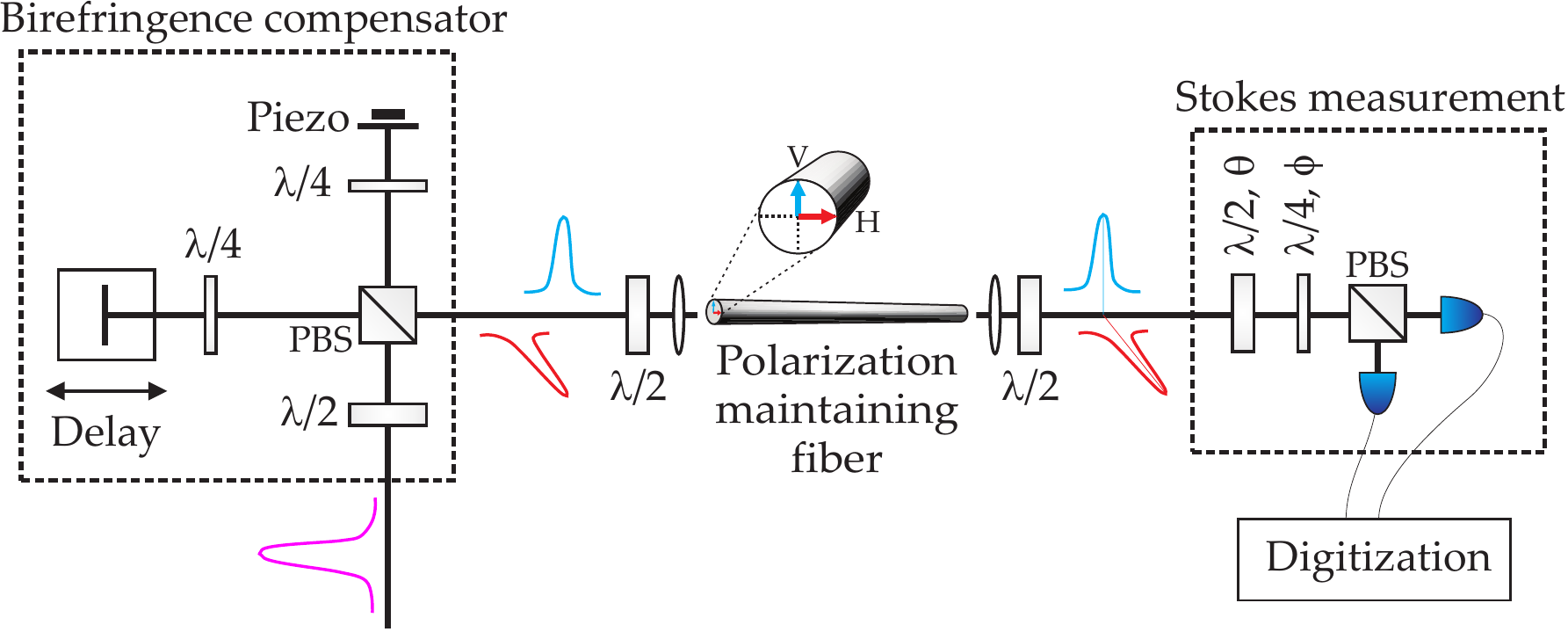}}
  \caption{Setup for efficient polarization squeezing generation and
    the corresponding Stokes measurement apparatus.}
\end{figure}

The quantum state is detected with a Stokes measurement, as sketched
in the previous section. The two detectors (InGaAS PIN photodiodes,
custom-made by Laser Components GmbH with 98~\% quantum efficiency at
DC) are balanced and have a sub-shot noise resolution at a frequency
range between 5~MHz and 30~MHz. Each detector has two separate
outputs: DC, providing the average values of the photocurrents, and
AC, providing the photocurrents amplified in radio-frequency (RF)
spectral range.  The RF currents of the photodetectors are mixed with
an electronic local oscillator at 12 MHz, amplified (FEMTO DHPVA-100),
and digitized by an analog/digital exit converter (Gage CompuScope
1610) at 10 Megasamples per second with a 16-bit resolution and 10
times oversampling.

The measurements are performed at a pulse energy of 93~pJ.  In the
dark plane a total squeezing of about 3.8~dB is observed. In the
orthogonal quadrature, the noise was enhanced by several tens of dB
due to Guided Acoustic Wave Brillouin Scattering
(GAWBS)~\cite{Shelby:1985ly,Shelby:1986ve,Elser:2006qf}. In the
direction of the classical excitation, the state is expected to be
shot-noise limited, since the Kerr effect only influences the phase 
and does not contribute to the photon number.
  
To perform the reconstruction, histograms of the Stokes variables are
recorded for different angles $(\theta, \phi)$.  This is done by
rotating the wave plates with motorized stages (OWIS DMT 40-D20-HSM)
and scanning one eighth of the Poincar\'e sphere in 8100 steps, a
measurement that took over eight hours. The unmeasured octants were
deduced from symmetry. For each setting of the wave plates, the
photocurrent noise of both detectors was simultaneously sampled $
0.5\times 10^6$ times. Noise statistics of the detectors difference
current were acquired in histograms with 751 bins. Additionally, the
optical intensity was recorded.

In figure~2 we show typical histograms at different angles on the
Poincar\'e sphere. As the widths largely vary from squeezing to
antisqueezing ranges, there are two plots in which the amplitude scale
differs in more than one order of magnitude.  The histograms labeled
1, 2 and 3 are measured in the dark plane. Tomogram 1 denotes the
angle of maximum squeezing, while 3 corresponds to the
antisqueezing. Tomogram 4 is at the classical mean value, where the
measured noise is almost shot-noise limited. Due to the high number of
samples, the measured histograms are smooth and, at the same time, the
number of bins makes it possible to resolve the large dynamical range
of amplitudes, so no data interpolation was needed. We also plot
histograms showing the electronic noise and the shot noise.

\begin{figure}
  \begin{center}
    \includegraphics[width=0.75\columnwidth]{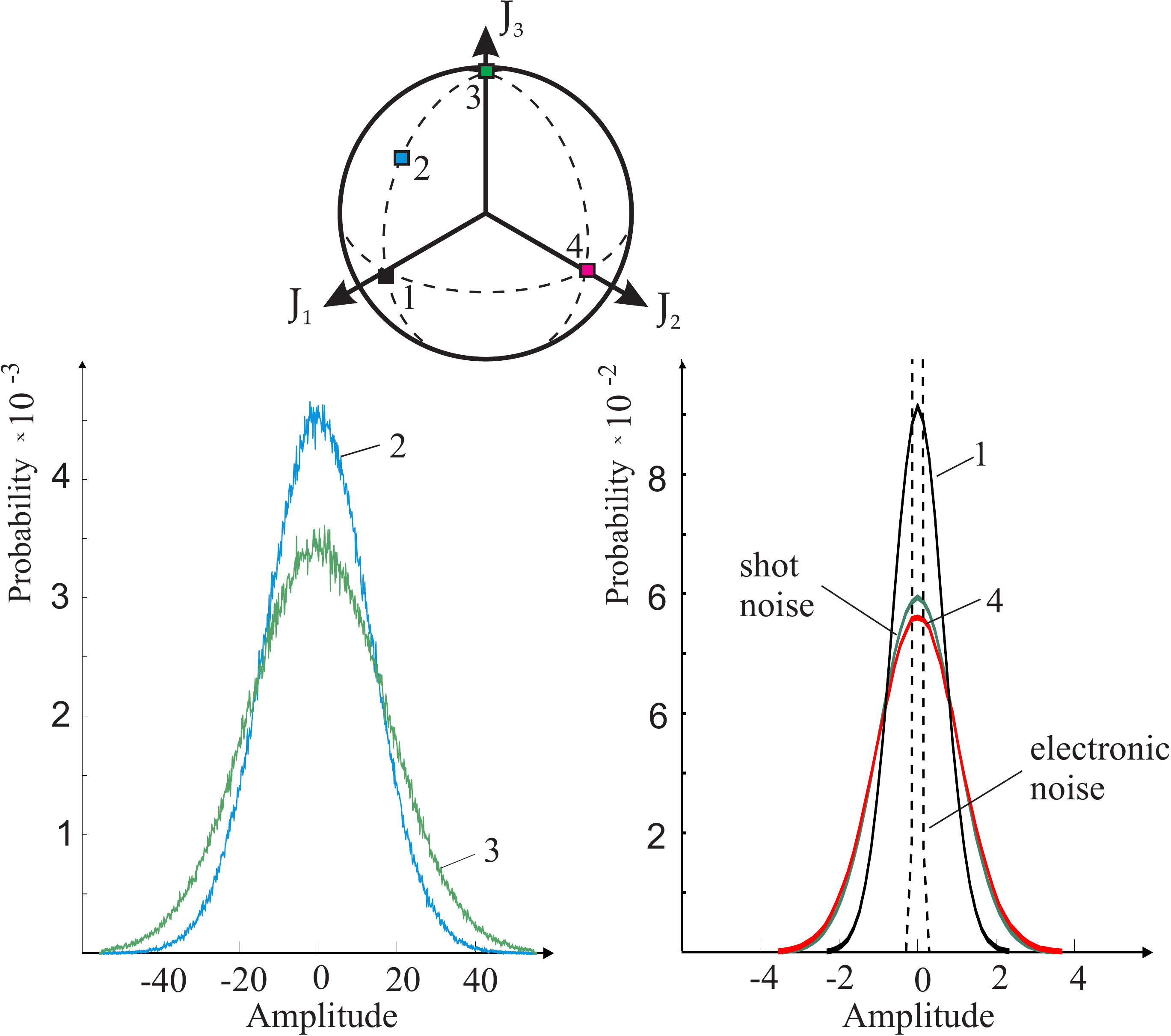}
    \caption{Measured histograms of the difference current of the two
      detectors for various measurement directions on the Poincar\'e
      sphere.. Note the different scales on both plots. Histograms 1,
      2, and 3 are in the dark plane, while histogram 4 is at the
      classical mean value. The histograms corresponding to electronic
      and shot noise are also shown.}
  \end{center}
\end{figure}

For all these histograms we have performed a Gaussianity check, using
the Kolmogorov-Smirnov and the $\chi^{2}$ tests, as well as the
Kullback-Leibler divergence\cite{Thode:2002cr}. We can conclude that
all the histogramas are Gaussian within confidence levels ranging from
95~\% to 98 \%.

\subsection{Experimental reconstruction}

As clearly expressed in (\ref{3DRT}), for high photon numbers the
tomography turns out to be equivalent to an inverse Radon transform of
the measured histograms.  In practice, this one-step 3D reconstruction
is very demanding in computational resources. Therefore, we divide the
process into two steps: first, a set of 2D projections is
reconstructed from the recorded histograms; subsequently, the Wigner
function is slice-wise generated from those projections (to which we
apply a Hamming filter to smooth the noise).  The symmetry of the
state is used as a prior information to reduce the range of measured
angles to an octant. This minimizes the systematic errors stemming
from imperfections of the polarization optics.

\begin{figure}
  \centering
  \includegraphics[width=0.95\columnwidth]{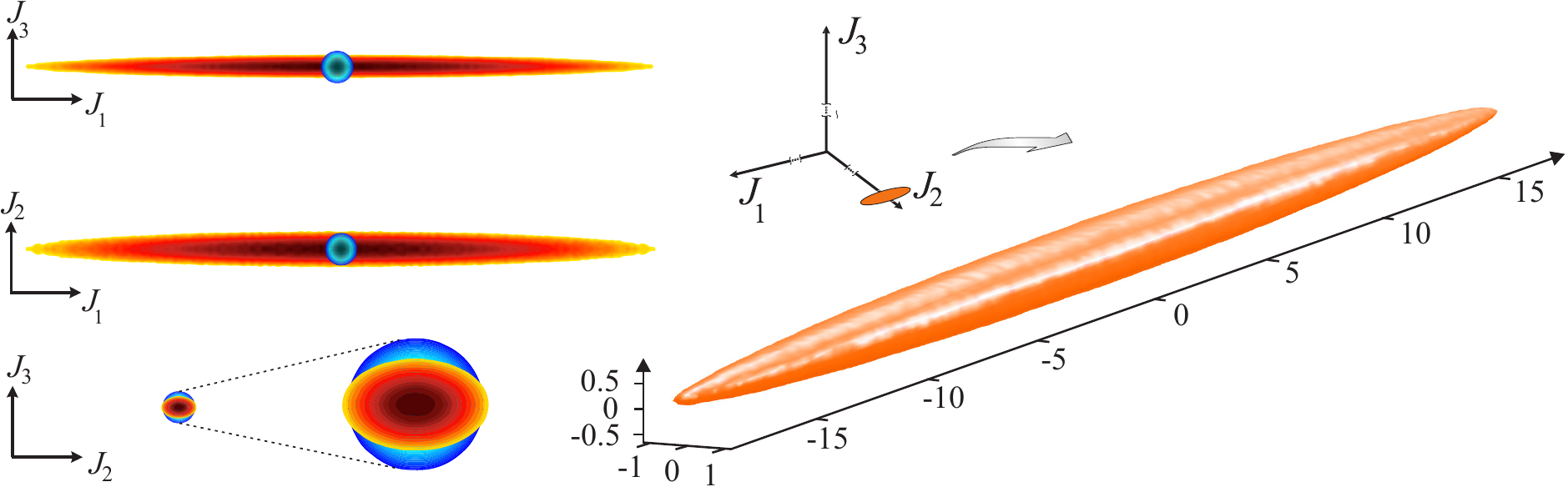}
  \caption{(Right) Isocontour surface of the level $1/e$ (from the
    maximum) of the Wigner function $W (J, \theta, \phi)$ for the
    polarization squeezed state generated in the setup of
    figure~1. (Left) Sections of the Wigner function through the three
    coordinate planes. In blue we show the isotropic section for a
    coherent state, which we use as unit for all the plots.}
\end{figure}

In figure~3 (right panel) we show the final result of the inverse
Radon transform for our polarization squeezed state.  More concretely,
we plot an isocontour surface of $W (J, \theta, \phi ) = \, $ constant
(with the constant being $1/e$ from the maximum) in the Poincar\'e
space having $J_1$, $J_2$, and $J_3$ as orthogonal axes. As coordinate
units we use the shot noise set by a coherent state. The ellipsoidal
shape of the state is clearly visible.  The center of the ellipsoid is
far away from the origin, since we have $10 \times 10^{11}$ photons
per measurement time (using 1.9~MHz resolution bandwidth).  The
antisqueezed direction of the ellipsoid is dominated by excess noise
stemming largely from GAWBS, as we have already mentioned.

In the left panel of figure~3 we sketch density plots of the
projections on the coordinate planes of the previous Wigner function
(including the particular case of a coherent state). The contours
agree with the 3.8 $\pm$ 0.3 dB squeezing that was directly measured
from the variances.  The projections on the planes $J_1$-$J_2$ and
$J_2$-$J_3$ show an additional spreading of the state in the $J_2$
direction caused by the imperfect polarization contrast in the
measurement setup that mixes some of the antisqueezing on the $J_2$
direction.

This Radon reconstruction requires a large set of measured data to get
a reasonably accurate representation of the state.  There are two main
reasons for this: integrals are approximated by finite sums (in our
case, we used 751 bins in 91 steps) and the kernel (\ref{kersing}) is
singular, so some \textit{ad hoc} filtering of the raw data is needed.
Acquiring such large data sets may be unwise, for it demands long
measurement times. Ensuring the proper stability of the setup is thus
essential and might be difficult depending on the quantum state
measured.
 
This limitation may be circumvented by adopting a
statistically-motivated method, such as the maximum likelihood
(ML)~\cite{ML:2004kl}.  In our case, the relation between the Wigner
function $W$ and the tomograms $w$ can be written as a system of
linear equations
\begin{equation}
  \label{maxlikeq}
  w_j=\sum_{j}c_{jk} \, W_k,
\end{equation}
where the subscripts in $W_{k}$ and $w_{j}$ is a shorthand notation
for the respectives coordinates. The coefficients $c_{jk}$ can be
interpreted as the overlap of the $j$th projector with the $k$th
volume element of the Wigner function and can be readily determined
from equations~(\ref{eq:WignSU2Agar}) and (\ref{cojo}).  The most
likely Wigner function is then found by minimizing the
Kullback-Leibler divergence between the normalized vectors of the
computed tomograms $w_j$ and recorded ones $\bar{w}_j$.  Technically,
this can be achieved by the iterative expectation-maximization
algorithm~\cite{Dempster:1977hc,Boyles:1983bs,Vardi:1993fv}
\begin{equation}
  \label{EMalgorithm}
  W_k^{(n+1)}=W_k^{(n)} \frac{\sum_j w_j}{(\sum_j \bar{w}_j) 
    (\sum_j c_{jk})} \sum_j\frac{\bar{w}_j}{w_j}  c_{jk} \, ,
\end{equation}
which converges monotonously to the ML estimate from any strictly
positive initial vector $W_k^{(0)}$.

\begin{figure}
  \centering
  \includegraphics[width=0.90\columnwidth]{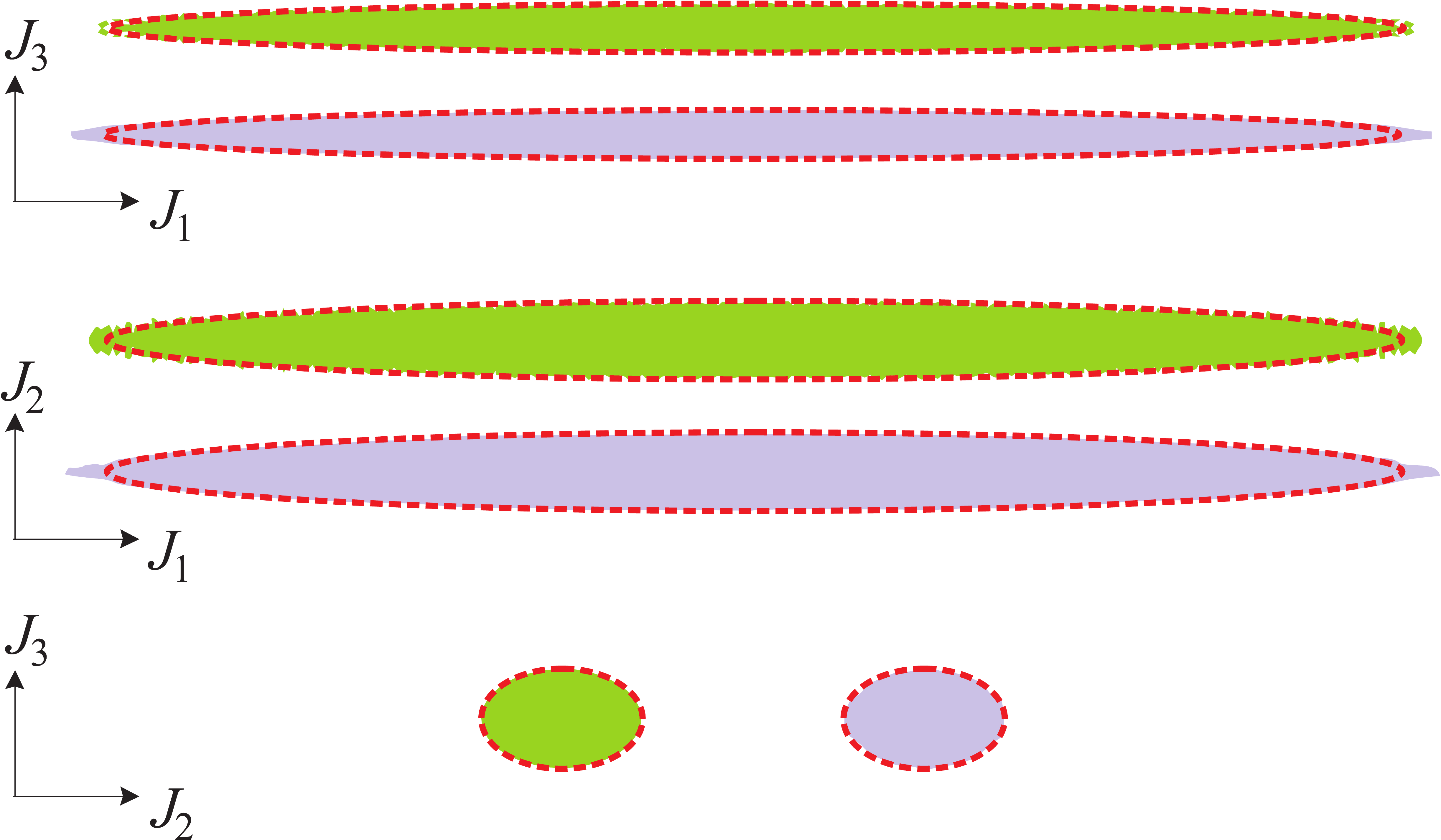}
  \caption{Sections of the isocontour of the Wigner function in
    figure~3 through the coordinate axes, for the three different
    reconstruction techniques used. In grey, direct Radon transform,
    in green ML method with nine settings for the angles of the wave
    plates and in dotted lines the results of a Gaussian ML
    approximation.}
\end{figure}

The significantly greater stability of the statistical inversion
allows us to get reconstructions of the same quality but from far
smaller data sets.  This is illustrated in figure~4, where we draw a
comparison between Radon and ML methods, although in the latter case
using only nine different settings of the angles $(\theta , \phi)$,
which amounts to reducing the measurements by two orders of
magnitude. In other words, the measuring time is shortened from eight
hours to less than five minutes! This result indicates that the
experimental characterization of considerably more complicated quantum
states with less symmetries should still be within the reach of the
present measurement setup.

\begin{figure}
  \includegraphics[width=\columnwidth]{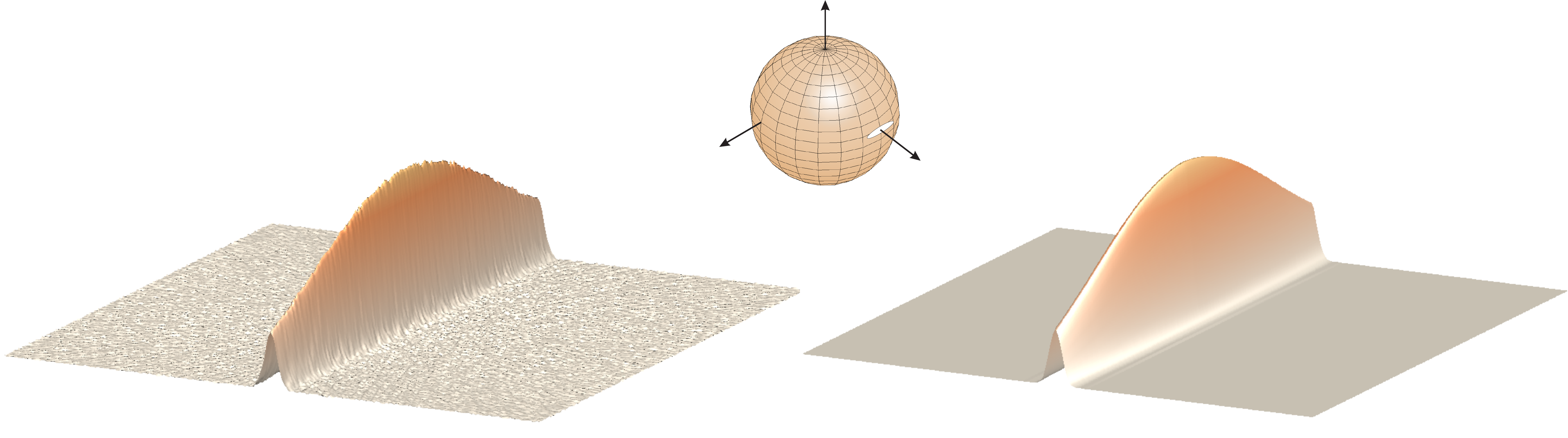}
  \caption{Dark plane reconstructions. Left panel: Reconstruction
    obtained by integrating the Wigner function shown in figure.~3 in
    the $J_2$ direction. Right panel: ML reconstruction from
    dark-plane histograms. Only nine settings of angles $\theta$ and
    $\phi$ were used for the ML tomography. \label{fig:darkplane}}
\end{figure}

As we have discussed in section~\ref{sec:darkplane}, the dark plane is
of special interest.  The theory shows that the reconstruction therein
can be obtained in two different ways: either by reconstructing the
dark mode directly from the histograms or by calculating projection of
the 3D Wigner function along the $J_2$ direction.  The two results are
compared in figure~\ref{fig:darkplane} and good agreement within the
experimental uncertainties is found. Since the Radon transform should
provide a plausible explanation for all the measured histograms, such
a comparison may serve as an independent test of the quality of the 3D
tomography.

Finally, the high confidence levels of the Gaussianity tests seems to
call for a Gaussian ML reconstruction. The Gaussianity is used as a
prior information about the signal, which helps to reduce drastically
the number of free parameters.  In this case, the Wigner function is
represented by the $3 \times 3$ covariance matrix $G$:
\begin{equation}
  W(\mathbf{n})\propto 
  \exp   \left  ( -\frac{1}{2} \mathbf{n}  G^{-1} \mathbf{n} \right ) \, , 
\end{equation}
and the calculated variances $\sigma_j=\mathbf{n}_j G \mathbf{n}_j$
are matched (in the ML sense) to the actually measured
variances~\cite{Rehacek:2009fu}.  Since $G$ must be positive
semidefinite, only six real parameters describe the measured system
and the problem is highly overdetermined, in consequence, the Gaussian
state can be obtained from a few histograms.  In principle, by
comparing Gaussian reconstructions based on different subsets of
measured data, various imperfections of the setup, such as
instabilities and biases, can be detected.

The matrix $G$ turns out to be
\begin{equation}
  \label{eq:Gaussiancov}
  G= \left (
    \begin{array}{ccc}
      3.0920 \times 10^{2} & -1.1931 & -2.0160 \\
      -1.1931 &   4.4485 \times 10^{-1} & -1.2926 \times 10^{-2} \\
      -2.0160 &  -1.2926 \times 10^{-2} & 1.1511 
    \end{array}
  \right ) \, ,
\end{equation}
which once diagonalized gives the principal variances 0.43962,
1.13853, and 309.22177 (in shot-noise units). This agrees well with
the standard and ML reconstructions, as can be also appreciated in
figure~4. The Gaussian reconstruction was done without assuming a
particular orientation or symmetry of the state with respect to the
Stokes coordinates. The covariance matrix suggests that the
misalignment of the principal axis is less than $0.5$ degrees within
the measurement errors, in accordance with the definition of angles
adopted in the experiment.

This Gaussian approach allows for a simple estimate of the errors:
just take the pseudoinversion of the measurement matrix as a linear
model and use the standard theory of error propagation.  The errors to
be propagated are actually the errors in the estimated variances for
each tomogram, which are found from the $\chi^{2}$ distribution.  In
addition, we can assume that the variances of different tomograms
are uncorrelated.

Taking a 97.5~\% confidence interval (which corresponds to three
standard deviations), the principal variances can be written as
\begin{equation}
  \label{eq:1}
  0.440 \pm 0.002 , 
  \quad
  1.139 \pm 0.001 ,
  \quad
  309.2 \pm 0.3 \, .
\end{equation}
Note that the relative errors in the two larger variances are roughly
the same ($\sim$ 0.1~\%), while for the smallest variance is four
times larger. This is a consequence of the experimental setup: the
smallest variance is directly revealed only in one of the recorded
projections used for the reconstruction.

\section{Concluding remarks}

In summary, we have presented a complete programme for the full
polarization tomography of quantum states. Using the SU(2) Wigner
function, we have provided an exact inversion formula in terms of the
histograms of a standard Stokes measurement and derived a simplified
version for very localized, high intensity states which turns out to
be an inverse Radon transform. As a test of the theory, the
reconstruction of an intense polarization squeezed state has been
performed.  A ML reconstruction algorithm has also been presented and
has been compared to the direct method, thereby yielding an excellent
agreement.  Of course, the technique can be readily used for any other
polarization state.

\ack

Two of us (ABK and LLSS) benefited from the friendship, wisdom, and
criticism of V~P~Karassiov. When finishing a long-due common review,
Valeri Pavlovich passed away unexpectedly. This paper is dedicated to
his memory.

We thank G~S~Agarwal, G~Bj\"{o}rk, S~M~Chumakov, H~de~Guise,
W~P~Schleich and K~B~Wolf for useful discussions.  We also acknowledge
A~Felipe for a careful check of the Gaussianity of our histograms.
Financial support from the EU FP7 (Grant Q-ESSENCE), the Spanish DGI
(Grants FIS2008-04356 and FIS2011-26786), the UCM-BSCH program (Grant
GR-920992), the Mexican CONACyT (Grant 106525), the Czech Ministry of
Education (Grant MSM6198959213), the Czech Ministry of Trade and
Industry (Grant FR-TI1/384) and the IGA of Palacky University (Grant
PRF-2011-005) is acknowledged.

\newpage


\providecommand{\newblock}{}

\end{document}